*Original Article*

# Role of ERP Modernization in Digital Transformation: PeopleSoft Insight

## Chandra Rawat

*Information Technology Services, Florida State University, Tallahassee, Florida, United States*

*Abstract - The role of Enterprise Resource Planning (ERP) systems with digital transformation strategies have become an important aspect of modern businesses to stay competitive in the fast-paced digital landscape. ERP modernization refers to the process of updating an organization's Enterprise Resource Planning (ERP) system to take advantage of the latest technological advancements and features. This article aims to provide insights into the role of ERP modernization in digital transformation and its impact on organizations. In particular, the article focuses on the PeopleSoft ERP system and its capabilities to support digital transformation initiatives. Peoplesoft, a widely used enterprise resource planning (ERP) system, has served organizations for over three decades. This article covers the functionalities of a modern ERP system; the benefits organizations can derive from implementing a modern ERP system, and the challenges organizations face in modernizing their ERP systems. This article also provides practical recommendations for overcoming these challenges and highlights the importance of considering factors such as scalability, security, and user experience while modernizing ERP systems.*

## 1. Introduction

Organizations must continuously adapt to new technologies and trends in the digital age to remain competitive and relevant in the market. Digital transformation has become a top priority for organizations across industries as they look to leverage digital technologies to improve their processes, increase efficiency, and stay ahead of the competition. ERP systems can be a key enabler of digital transformation by providing organizations with a single, integrated platform for managing key business processes. With an ERP system in place, organizations can automate and streamline a range of tasks, from financial and supply chain management to human capital and customer relationship management.

PeopleSoft is one of the most well-known and widely used ERP systems and has served organizations for over two decades. The system was originally developed to help organizations manage and automate their core business processes. It has since evolved to offer a wide range of functionalities, including financial and supply chain management and human capital management. One of the key strengths of PeopleSoft is its flexibility and ability to adapt to changing business needs. With PeopleSoft, organizations can configure and customize the system to meet their specific requirements, allowing them to improve processes and increase efficiency over time. However, as technology has advanced and business needs have changed, many organizations are finding that their Peoplesoft systems are becoming outdated and needing modernization. This has led to a growing trend of organizations seeking to modernize their Peoplesoft systems to improve functionality, user experience, and overall business outcomes.

The following diagram below illustrates the relationship between the components of the PeopleSoft Internet Architecture [1].

## 2. Reasons for ERP Modernization

There are several reasons why organizations should modernize their ERP systems and accelerate digital transformation. Some of the most common reasons include [2][14][15].

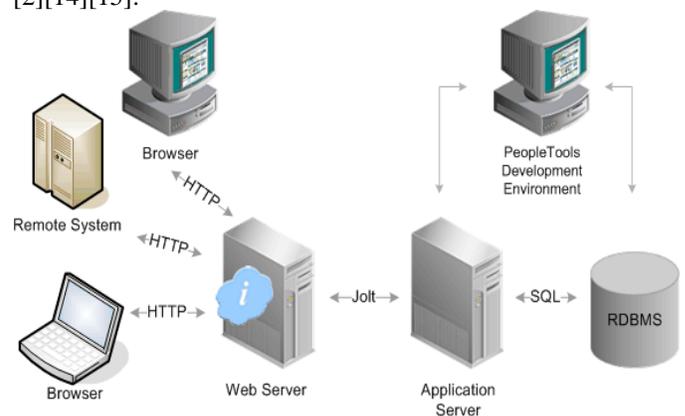

**Fig. 1 PeopleSoft Internet Architecture**

### 2.1. Outdated Technology

Peoplesoft ERP systems in their older versions are used by many businesses even though these systems are no longer supported by Oracle or compatible with modern hardware or software.

### 2.2. Limited Functionality

Old versions of ERP systems may not have the functionality required to meet the evolving needs of organizations.

### 2.3. Poor User Experience

Users may find the ERP system user interface outdated or difficult to use, leading to decreased productivity and user adoption.

### 2.4. Integration Challenges

Integrating outdated ERP with other systems and technologies can be difficult and time-consuming, leading to increased costs and decreased efficiency.

### 2.5. Data Security Vulnerabilities

ERP systems from various IT infrastructure hardware and software have been assembled over the years. Due to this, the systems have several unknown vulnerabilities that can certainly be taken advantage of by hackers. Many high-profile data breaches have occurred in recent years, many of which have been traced back to legacy systems' outdated data security architecture.

### 2.6. Business Challenges

Obsolete ERP systems hurt productivity as skilled IT employees who should be creating value are charged to maintain the system. Additionally, it is difficult to meet workforce expectations, such as providing employees with a mobile or smart device interface.

### 2.7. Support and Maintenance Costs

Old versions of ERP system Support and maintenance can cost an organization thousands of dollars every year, and that cost may overtake the benefits provided by the system.

## 3. Characteristics of a Modern ERP and PeopleSoft's Latest Capabilities

The following are some of the key characteristics of modern ERP system materials.

### 3.1. User Interface

Modern ERP web pages should be Responsive, allowing them to change their layout and content to any device used to access them. This ensures the website is optimized for the greatest possible user experience on all devices, including desktop computers, tablets, and smartphones. Back in 2014, Oracle introduced Fluid User Interface (UI) technology for PeopleSoft. The fluid user interface enables access to PeopleSoft applications on several form factors, including smartphones, tablets, and desktops/laptops. Regardless of screen size, fluid applications provide a consistent user experience on a multitude of devices [27][28].

### 3.2. Information Security

Information security refers to the steps taken to prevent unauthorized access, use, disclosure, interruption, alteration, or destruction of sensitive data, systems, and networks. Information security is essential for enterprises, organizations, and individuals to secure the confidentiality, integrity, and availability of their data [24][25][27]. PeopleSoft ERP system includes a robust security framework to help protect sensitive data. Role-based security in PeopleSoft grants accesses based on job duties [3].

PeopleSoft prevents unauthorized access through traditional user IDs and passwords. It also offers multi-factor authentication and integrates with third-party authentication systems such as Microsoft Active Directory and Okta. The most recent version of PeopleSoft encrypts critical data, including Bank Account numbers and personal information, to safeguard data privacy.

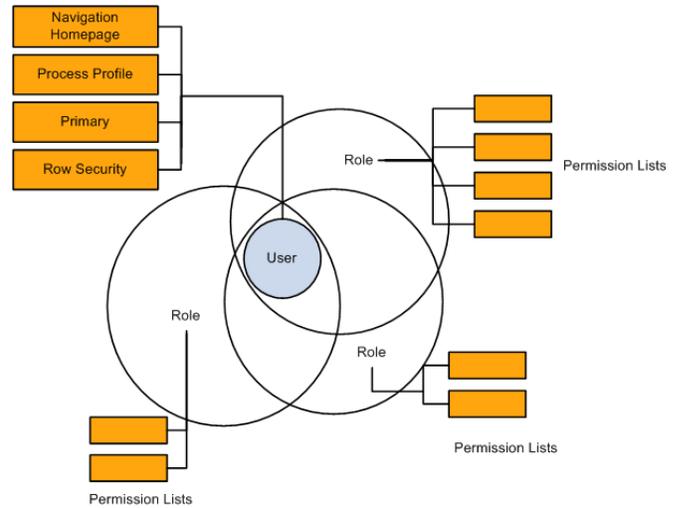

**Fig. 2 PeopleSoft Role-Based Security Framework**

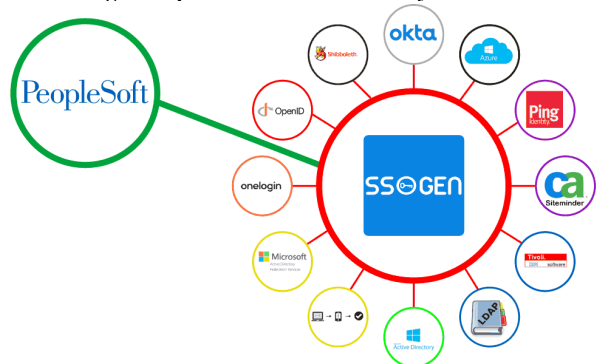

**Fig. 3 PeopleSoft Top Third-Party Authentication Systems**

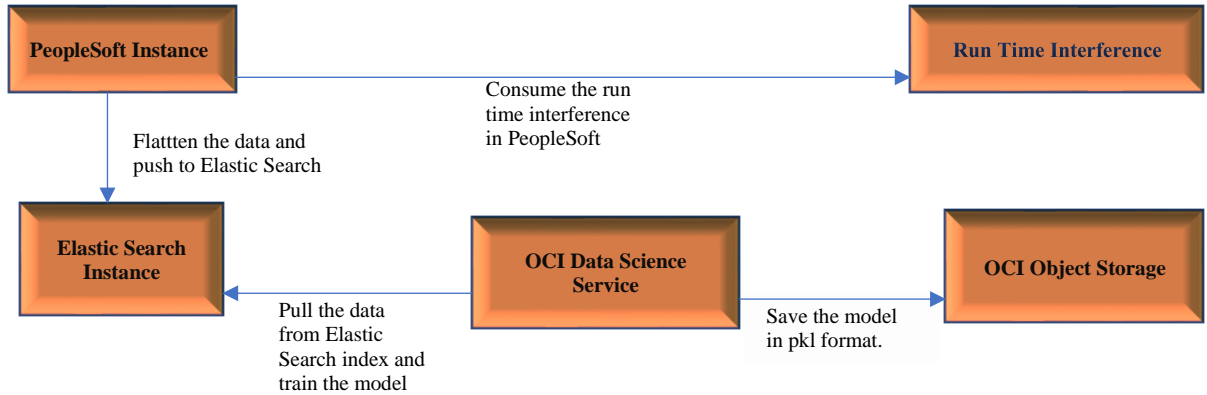

**Fig. 4 OCI Data Science Architecture for PeopleSoft**

### 3.3. Artificial Intelligence

Artificial Intelligence (AI) is increasingly being integrated into enterprise resource planning (ERP) systems to enhance their capabilities and improve efficiency [17][18]. AI technology has the potential to enhance many aspects of ERP systems and provide businesses with new insights and improved decision-making capabilities. Here are some of the ways that AI can be used in ERP systems. Oracle PeopleSoft lacks artificial intelligence (AI) capabilities. However, Oracle is always working to improve the artificial intelligence capabilities of PeopleSoft and now offers OCI Data Science, a new service that is part of Oracle Cloud Infrastructure (OCI). It offers Data Scientists a platform on which they can develop, train, and administer Machine Learning Models by leveraging the Python Jupyter Notebook [4].

PeopleSoft also released a chatbot in 2019. The Chatbot Integration Framework for PeopleSoft is a collection of resources Oracle provides through PeopleTools 8.57.07 (and later versions) and PeopleSoft Update Images. Its purpose is to enable the use of digital assistants within PeopleSoft applications by connecting them to the Oracle Digital Assistant (ODA) service. Absence Assistant, Requisition Assistant, and Company Directory are examples of Oracle-delivered chatbots. The Oracle PeopleSoft team created the Chatbot Integration Framework in PeopleTools to make integrating Chatbots with PeopleSoft applications simple and safe. [5].

### 3.3.1. Low-code and no-code

IT tools that allow for the rapid development of business applications are important in today's dynamic business climate. Both low-code and no-code platforms can help organizations speed up their application development process, reduce the time and cost of development, and allow for more rapid prototyping and iteration. The main difference between the two is the amount of coding required to build an application. Modern ERP should have the functionality to build an application quickly. Additionally, it should have robust configuration capability to change the business processes without much need for coding [19][20].

Peoplesoft initially lacked these capabilities, and clients had to make extensive customizations to satisfy their business requirements, resulting in lengthy, costly, and difficult maintenance for Peoplesoft's business suit. However, in the last few years, PeopleSoft has developed several tools to remove customization and provide better configuration setups. This area still requires a great deal of work. Users of PeopleSoft can now utilize an event mapping framework that permits the injection of custom code into component and component field events without modifying the original objects. The capability makes new PUM upgrades simpler, less expensive, and faster, allowing customers to satisfy their specific business requirements. Another tool provided by PeopleSoft is page composer, which allows users to change the design layout of certain PeopleSoft pages simply by dragging and dropping without any coding.

### 3.3.2. Analytics and Reporting

Modern ERP technology uses data visualization, reporting dashboards, and key performance indicators (KPIs) to enhance visibility, performance monitoring, and decision-making. Modern ERP systems should offer interactive and intuitive data visualization features to help organizations quickly understand their data. It should provide reporting dashboards to track performance and goals. Dashboards give organizations real-time data and insights to identify and address issues. Key Performance Indicators (KPIs) help organizations track progress toward targets. KPIs track performance and project success in current ERP systems. Overall, these can help organizations make informed decisions and gain better visibility into their operations.

Oracle recently integrated Kibana with PeopleSoft. Kibana for PeopleSoft offers extensive analytics, visualisation, and performance advantages, enabling better decision-making and understanding of organizational data. The diagram below represents the PeopleSoft server architecture integrated into Elasticsearch, Kibana, and Logstash [6] [7] [8].

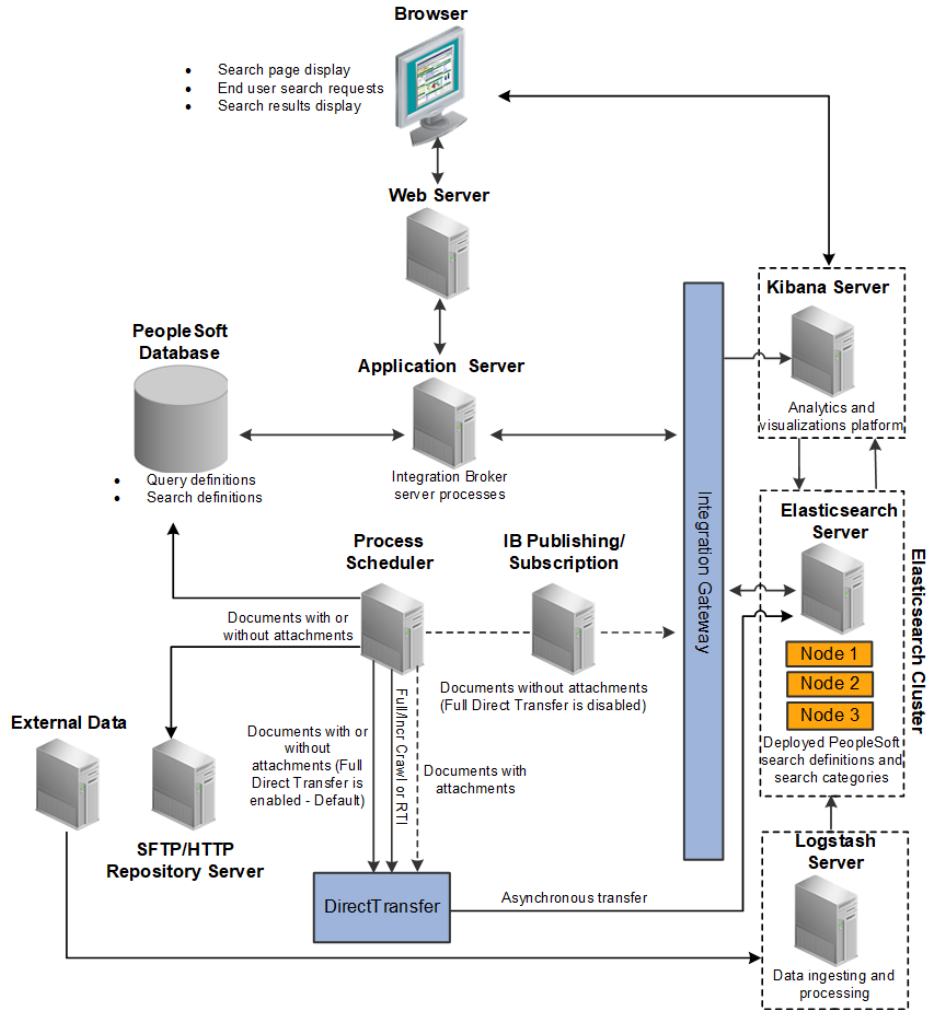

**Fig. 5 PeopleSoft architecture connected to Elasticsearch Kibana, and Logstash**

*Automation*

Modern ERP systems automate and streamline processes, reducing manual errors and increasing efficiency. This can help organizations save time and resources, allowing them to focus on other important tasks [22][23].

Automation is an important component of PeopleSoft, allowing organizations to streamline processes and increase efficiency. Some examples of PeopleSoft automation features include:

*a) Workflow Automation*

PeopleSoft includes a workflow engine that automates the flow of tasks, approvals, and employee communications. Automated approval processes such as expenses, purchase orders, and time-sheet requests are great examples of PeopleSoft Automation. This can help organizations save time and reduce manual errors.

*b) Self-service Portals*

PeopleSoft includes self-service portals that allow employees to perform tasks such as updating their personal information, submitting time off requests, and accessing their pay stubs. This can help organizations reduce the workload of HR and administrative staff.

*c) Data Integration*

PeopleSoft integrates with other systems and technologies, allowing organizations to automate processes and reduce manual errors. This can help organizations save time and increase efficiency. PeopleSoft supports both RESTful Webservice, HTTP, and SOAP-based web services [9].

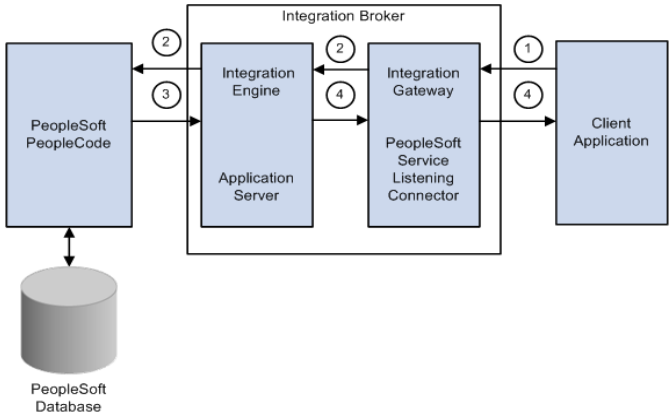

**Fig. 6 PeopleSoft Web services integration**

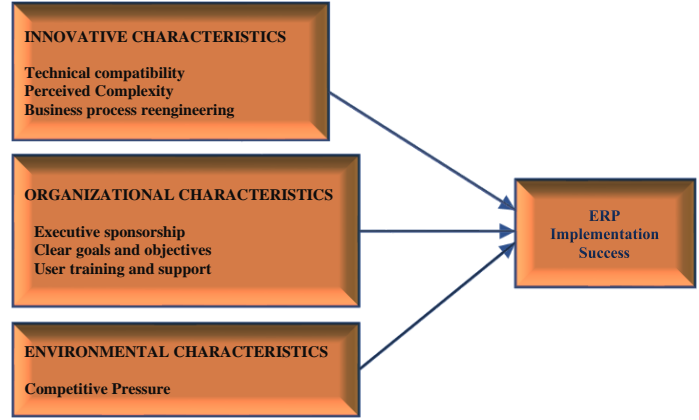

**Fig. 7 ERP Research Model [2]**

*d) Automated Reporting*

PeopleSoft includes a reporting engine that generates reports based on data from the system. This can help organizations gain better visibility into their operations and make informed decisions.

*Cloud-based Software*

ERP solutions based on the cloud are gaining popularity for various reasons.

With a cloud-based ERP system, employees may use the system from anywhere with an internet connection, which is very advantageous for businesses with remote workers or several locations compared to on-premises systems; cloud-based ERP solutions often offer cheaper initial expenses and lower ongoing maintenance costs. Cloud-based ERP systems are automatically upgraded by the provider, eliminating the need for enterprises to update the software manually. Unlike on-premises systems, cloud-based ERP solutions may be modified and adapted to match the unique requirements of any firm.

PeopleSoft's Cloud-based ERP systems offer organizations the benefits of the traditional PeopleSoft ERP solution with the added benefits of the cloud [10] [11].

## 4. Results and Discussion

### 4.1. Critical Factors for a Successful *ERP Modernization or Upgrade*

Several industries have adopted ERP systems over the last decade. ERP systems have transformed corporate computing by allowing streamlined and real-time planning, manufacturing, and customer response. However, many organizations still depend on legacy Software systems or an old ERP version. Many organizations have complained about unsuccessful ERP modernization/ upgrades.

The following critical components, which also apply to ERP system modernization, have been identified by researchers as crucial factors that may contribute to the successful deployment of an ERP system [2][13][29].

The following are critical success factors for an ERP modernization or upgrade.

### 4.1.1. Technical Compatibility

Technical compatibility implies the compatibility of an ERP with existing systems, including infrastructure. Certain software will likely be retained in an ERP modernization and must be integrated with the modern system. The level of ERP system compatibility with retained technical systems has a favorable association with modernization success.

### 4.1.2. Business Process Reengineering

Most analysts believe customization contributes to costly and lengthy implementations. Thus, organizations should maintain the ERP suit as it is as much as possible and redefine their business processes to the best practices. The degree of reengineering to an ERP system's best practices strongly correlates with modernization success.

### 4.1.3. Change Management

The ERP modernization project will likely impact many parts of the organization, and change management strategies should be in place to ensure a smooth transition.

### 4.1.4. Continuous Improvement

An ERP modernization project is not a one-time event but an ongoing continuous improvement process. The system should be regularly reviewed and improved to ensure it continues to meet the organisation's needs.

### 4.1.5. Organizational Characteristics

According to this ERP implementation study, the culture of the business is crucial to implementation success and applies to modernization as well. Senior leadership's encouragement, training, and agreement with organizational goals are important cultural factors in shaping success [2].

### *4.1.6. Executive Sponsorship*

The support and commitment from the highest level of leadership are crucial for ensuring that the ERP modernization project is a priority for the organization.

### *4.1.7. Clear Goals and Objectives*

All stakeholders in the ERP modernization project need to clearly understand what the organization is trying to accomplish and how the new ERP system will support these objectives. Having clear goals and objectives helps to ensure that everyone is working towards the same result and that the project stays on track. This can include specific outcomes such as improved efficiency, better data management, increased collaboration, or more streamlined business processes. Additionally, having clear goals and objectives can help the organization prioritize its efforts and allocate resources appropriately.

### *4.1.8. User Training and Support*

It is essential to provide users of the new system with adequate training and support to ensure that they can use the system efficiently and that its adoption is successful.

### *4.1.9. Environmental Characteristics*

The study says that one of the most often cited reasons for implementing an ERP is competitive pressure [30].

Competitive pressure: Effective business plans are driven by the desire to build and maintain a strategic edge in the marketplace. Competitive pressure to modernize an ERP suite can positively impact implementation success.

## 5. Conclusion

ERP modernization is an essential aspect of digital transformation for organizations. Modern ERP systems can accelerate manual operations' automation, allowing more time and money to be spent on other key projects. Modern ERP systems can aggregate data for easier access, management, and analysis. This helps with data-driven decision-making. The latest ERP system enhancements eliminate manual procedures, reduce errors, and speed up business operations. Modern ERP systems can assist businesses in improving customer service using AI and machine learning tools such as Chatbots and predictive analysis. Additionally, modern ERP systems can provide real-time information on customer orders and are highly adaptable to new and changing business practices.

The PeopleSoft ERP system is an excellent option for businesses seeking to modernize their ERP systems, as it offers a variety of ERP solutions to fulfill the unique requirements of various businesses. PeopleSoft Financials suit helps organizations to manage accounts payable, receivable, and general ledger. PeopleSoft Human Capital Management (HCM) suite manages personnel records, payroll, benefits, and talent. PeopleSoft Supply Chain Management (SCM) suite helps companies manage supply chain operations, including procurement, inventory management, and order management. PeopleSoft Campus Solutions suite helps universities handle admissions, enrollment, and financial assistance.

Additionally, organizations can choose to implement one or more PeopleSoft suites to create a comprehensive ERP solution that meets all their needs. PeopleSoft has recently been at the forefront of innovation, focusing on cloud-based deployment and mobile-friendly interfaces. This has made it easier for organizations to access the system from anywhere, at any time, and has helped further improve business processes' efficiency and speed.

However, the organization needs to be aware of the difficulties that come along with modernizing ERP systems, such as the absence of expertise within the organization and the high cost of implementation. Organizations can ensure they make the most of their investment in digital transformation initiatives by considering factors such as scalability, security, and user experience while modernizing ERP systems.